\documentclass[a4paper,11pt]{article}
\pdfoutput=1 

\usepackage{jinstpub} 

\title{\boldmath First results of  Resistive-Plate Well (RPWELL) detector operation at 163 K}

\author[a,b,1]{A. Roy,\note{Corresponding author.}}
\author[c]{M. Morales,}
\author[d]{I. Israelashvili,}
\author[a]{ A. Breskin,}
\author[a]{S. Bressler,}
\author[c]{D. Gonzalez-Diaz,} 
\author[e]{\\C. Pecharrom\'an,}
\author[a]{S. Shchemelinin,}
\author[a]{D. Vartsky}
\author[b]{and L. Arazi.}

\affiliation[a]{Department of Particle Physics and Astrophysics, Weizmann Institute of Science, Rehovot 7610001, Israel}
\affiliation[b]{Nuclear Engineering Unit, Faculty of Engineering Sciences, Ben-Gurion University of the Negev, Beer-Sheva 8410501, Israel}
\affiliation[c]{Instituto Galego de Fisica de Altas Enerxias, 15782 Santiago de Compostela, Galicia, Spain}
\affiliation[d]{Physics Department, Nuclear Research Centre Negev, Beer-Sheva 9001, Israel}
\affiliation[e]{Instituto de Ciencia de Materiales de Madrid-CSIC, C/ Sor Juana In\'es de la Cruz, 3, Cantoblanco, 28049 Madrid, Spain}

\emailAdd{arindam.roy@weizmann.ac.il}

\abstract{We present for the first time, discharge-free operation at cryogenic conditions of a Resistive-Plate WELL (RPWELL) detector. It is a single-sided Thick Gaseous Electron Multiplier (THGEM) coupled to a readout anode via a plate of high bulk resistivity. The results of single- and double-stage RPWELL detectors operated in stable conditions in Ne/5$\%$CH$_{4}$ at 163 K are summarized. The RPWELL comprised a ferric-based (Fe$^{3+}$) ceramic composite ("Fe-ceramic") as the resistive plate, of volume resistivity $\sim$$10^{11}$ $\Omega$$\cdot$cm at this temperature. Gains of $\sim$$10^{4}$  and  $\sim$$10^{5}$ were reached with the single-stage RPWELL, with 6 keV X-rays and single UV-photons, respectively. The double-stage detector, a THGEM followed by the RPWELL, reached gains $\sim$$10^{5}$ and $\sim$$10^{6}$ with X-rays and single UV-photons, respectively. The results were obtained with and without a CsI photocathode on the first multiplying element. Potential applications at these cryogenic conditions are discussed.}

\keywords{Micropattern gaseous detectors (MSGC, GEM, THGEM, RETHGEM, MHSP, MICROPIC, MICROMEGAS, InGrid, etc), Resistive-plate chambers, Electron multipliers (gas)}

\begin{document}
\maketitle
\flushbottom

\section{Introduction}
\label{sec:intro}

\begin{figure}[!h]
\centering 
\includegraphics[width=.75\textwidth]{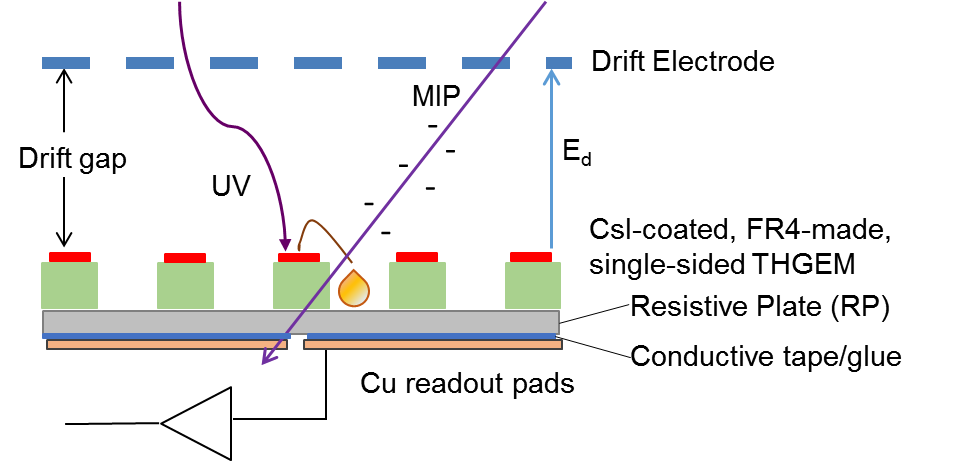}
\caption{\label{fig:i} A schematic diagram of the single-stage Resistive-Plate WELL detector. The single-sided THGEM is coupled to a readout anode via a resistive material ($\rho$ $\sim$$10^{9}$ - $10^{12}$ $\Omega$$\cdot$cm). Induced signals are read out capacitively on the segmented anode underneath. An optional photocathode (CsI, shown here) deposited on the THGEM electrode, makes the detector sensitive to UV light.}
\end{figure}

The present work aims at developing high-gain robust detection elements capable of operation at cryogenic temperatures -- such as in noble-liquid detectors. The ones investigated here are Resistive-Plate WELL (RPWELL) detectors \cite{1,2} in single- and two-stage configurations. The RPWELL is a robust, single-stage gas avalanche multiplier. It consists of a single-sided Thick Gaseous Electron Multiplier (THGEM) \cite{3,4,5} electrode coupled to a readout anode via a resistive plate (Figure \ref{fig:i}). Ionization electrons induced by X-rays or UV-induced photoelectrons from a photocathode deposited on the top surface (e.g. CsI) are collected into the THGEM holes where they undergo avalanche multiplication. Signals are induced capacitively through the resistive plate onto a patterned readout anode,  in direct contact with the resistive plate. Similar to other Micro Pattern Gaseous Detectors (MPGDs) using resistive elements \cite{6,7,8,9,10,11}, the RPWELL concept aims at quenching occasional electrical discharges occurring when the avalanche size exceeds, within a single hole, the theoretical Raether limit \cite{12}. Such discharges may cause significant damage to the electrodes and readout electronics along with substantial dead-times and gain variations due to charging up of the detector elements \cite {13,14,15,16}. In the configuration shown in Figure \ref{fig:i}, unlike other multipliers with resistive electrodes (Resistive Micromegas \cite{6}, Resistive WELL (RWELL) \cite{7}, Segmented Resistive WELL (sRWELL) \cite{8,9} and Micro Resistive WELL ($\mu$-RWELL) \cite{10}), the avalanche electrons are transported perpendicularly through the resistive layer, rather than spreading sideways. This minimizes the cross-talk between adjacent readout elements (pads or strips) and is particularly important in "digital" detectors involving "pad-counting", like the sampling elements developed for Digital (or semi-digital) Hadron Calorimetry (DHCAL; sDHCAL) \cite{17,18,19,20}. Discharge quenching in an RPWELL is governed essentially by the plate{\textquotesingle}s bulk resistivity ($\rho$); its optimal value should range within $\sim$$10^{9}$ - $10^{12}$ $\Omega$$\cdot$cm \cite{1,21,22}. Lower resistivity values fail to reduce the discharge energy to a sufficient level, while higher resistivity values lead to charging up of the detector-electrode materials, resulting in rate-dependent gain fluctuations \cite{1,15,16}. Due to the high resistivity of the electrodes, the discharges are confined within a small geometrical area of $\sim$ a few mm$^{2}$ \cite {23,24} and induce a local voltage drop. The resistance (R) offered by the resistive plate to the flow of charge along with the effective electrode capacitance (C), define an RC circuit, characterized by a time constant $\tau =  RC = \rho\epsilon$, where $\rho$ and $\epsilon$ are the bulk resistivity of the resistive plate and the dielectric permittivity of the electrode material, respectively. The time constant $\tau$, determines the recovery time of the system after a discharge, i.e. the time taken by the power supply to recharge the electrode to the applied voltage. Typical recovery time scales for resistive plates having resistivity in the aforementioned range are $\sim$$10^{-1}$ - $10^{2}$ ms, considerably longer than the typical discharge time scale of several nanoseconds \cite{24}. Therefore, in the presence of the resistive plate, the electrodes behave effectively as insulators and discharges are quenched.

Detailed investigations of RPWELL detectors in the laboratory \cite{1,15} and with particle beams \cite{25,26} yielded two optimal materials for room-temperature (RT) applications --- Semitron ESD 225 (Quadrant Plastics, USA, $\rho$ $\sim$$10^{9}$ $\Omega$$\cdot$cm) and Low Resistive Silicate (LRS) Glass (Tsinghua, China, $\rho$ $\sim$$10^{10}$ $\Omega$$\cdot$cm) \cite{27} (both values at RT). With the aim of developing robust sampling elements for (s)DHCAL in future particle-physics experiments, studies have been carried out with detector prototypes reaching sizes of 480 $\times$ 480 mm$^{2}$ \cite{20}. Room-temperature tests with RPWELL detectors have demonstrated discharge-free operation at high gains ($>$$10^{4}$) \cite{25,26,28}, with high detection efficiency of minimum ionizing particles (MIPs) \cite{28} and sub-millimetric position resolution \cite{29}, at rates $\sim$$10^{4}$ Hz/cm$^{2}$ \cite{28}, also in the presence of highly-ionizing background.

An important potential application for RPWELL-based detectors is in liquid xenon (LXe) and liquid argon (LAr) detectors. Examples are dark-matter searches \cite{30,31}, neutrino physics \cite{32}, gamma-radiation and fast-neutron imaging in homeland security \cite{33}. Noble-liquid detectors could make use of RPWELL elements in different configurations, recording either UV-photons or ionization charges. One example is their incorporation as a last amplification stage in a cryogenic gaseous photomultiplier (GPM), e.g. with a CsI photocathode \cite{34,35}, to detect and localize primary scintillation and/or electroluminescence photons. GPMs offer high spatial resolution (unlike conventional PMTs) and by far lower dark current compared to silicon photomultipliers (SiPMs) -- a vital requirement in detectors searching for rare, low-energy signals, such as in dark-matter experiments. Compared to current cryogenic CsI-GPM{\textquotesingle}s, with cascaded THGEM elements (investigated at temperatures $\sim$180 K; coupled to LXe) \cite{35}, a last-stage RPWELL electrode would provide superior stability, as demonstrated below. 

Another potential application would be the incorporation of RPWELL-elements in dual-phase LXe and LAr detectors \cite{36}. The most appropriate example is the current dual-phase LAr TPC developed as an option for the DUNE neutrino experiment \cite{37}; the large 6$\times$6$\times$6 m$^{3}$ DUNE prototype in the CERN-WA105 demonstrator currently relies on Large Electron Multiplier (LEM) charge-multiplying elements \cite{36,38,39}. The stable gain of the latter in MIPs tracking in LAr is $\sim$20 -- limited by occasional discharges \cite{39}. Here, effective spark-quenching by RPWELL elements is expected to yield higher charge gains, thus reducing the detection thresholds.

The aim of the present study is to demonstrate, for the first time, the feasibility of RPWELL-detector operation at cryogenic temperatures. As noted above, discharge quenching requires a bulk resistivity in the range of  $\sim$$10^{9}$ - $10^{12}$ $\Omega$$\cdot$cm. The main challenge has been to find an appropriate material in that resistivity range -- at cryogenic temperatures. Unfortunately, the vast majority of resistive materials display an Arrhenius behavior \cite{40}, i.e. an exponential increase of the resistivity with decreasing temperatures, turning them effectively into insulators at LXe and LAr temperatures\footnote {The triple point of Xe is 161.4 K, 0.82 bar and of Ar is 83.8 K, 0.69 bar. LXe detectors are typically operated over a pressure range of 1-2.5 bar, corresponding to 165 K-183 K. LAr detectors are typically operated at 1 bar, corresponding to 87 K.}\cite{41}. As discussed below, materials investigated so far whose resistivity does not follow this behavior, either have too low temperature-independent resistivity ($\rho$) $\sim$$10^{6}$ - $10^{8}$ $\Omega$$\cdot$cm, or suffer from severe non-uniformities. 

We report here on the first successful  operation of an RPWELL detector having a resistive plate made of ferric-based (Fe$^{3+}$) ceramic composite ("Fe-ceramic") \cite{42} at LXe temperature; at 163K, close to the triple point of Xe, with the resistivity tuned to a suitable value at this temperature. The results of the search for appropriate resistive materials are briefly summarized. We describe the properties of small Fe-ceramic RPWELL detector prototypes in single-stage RPWELL and double-stage THGEM+RPWELL configurations, in Ne/5$\%$CH$_{4}$. Results are presented at RT and at 163 K, with soft X-rays and UV-photons: among them, the maximum achievable stable gain, energy resolution and discharge behavior.

\section{Resistivity Measurements}

Resistivity measurement of different candidate materials were performed from RT down to cryogenic temperatures, aiming at  $\rho$ $\sim$$10^{9}$ - $10^{12}$  $\Omega$$\cdot$cm at LXe and LAr temperature. The materials investigated were $\sim$ 3 $\times$ 3 cm$^2$ square plates, with a thickness of 0.4 - 2 mm.  They were painted with conductive Ag paint (Leitsilber 200 Silver Paint, Ted Pella Inc., USA) on both faces and pressed between 4 mm thick PTFE blocks having spring-loaded pin contacts. The pins ensured proper contact for the high voltage bias and the current readout. The plates were biased to a potential {\it V} with a CAEN N471A power supply and the current {\it I} through the bulk of the resistive plate was monitored with a Keithley 610 CR Pico-ammeter. The volume resistivity was then calculated as:

\begin{equation}
\label{eq:y:1}
\rho =  R \frac{A}{l}
\end{equation}

\noindent where {\it R} is the resistance, ${\it A}$ is the area of the electrode and ${\it l}$ is the plate thickness. 
The resistance {\it R} was calculated by a linear fit to the I-V curve, for materials exhibiting Ohmic behavior. 
Measurements were performed in a vacuum chamber (held at $10^{-5}$ mbar) immersed in a LN$_{2}$ bath; resistivity values of the samples were measured under repeated cooling and warming up cycles. The sample temperature was monitored by a thermocouple placed in direct contact with it. 

\begin{figure}[!ht]
\centering
\includegraphics[width=.75\textwidth]{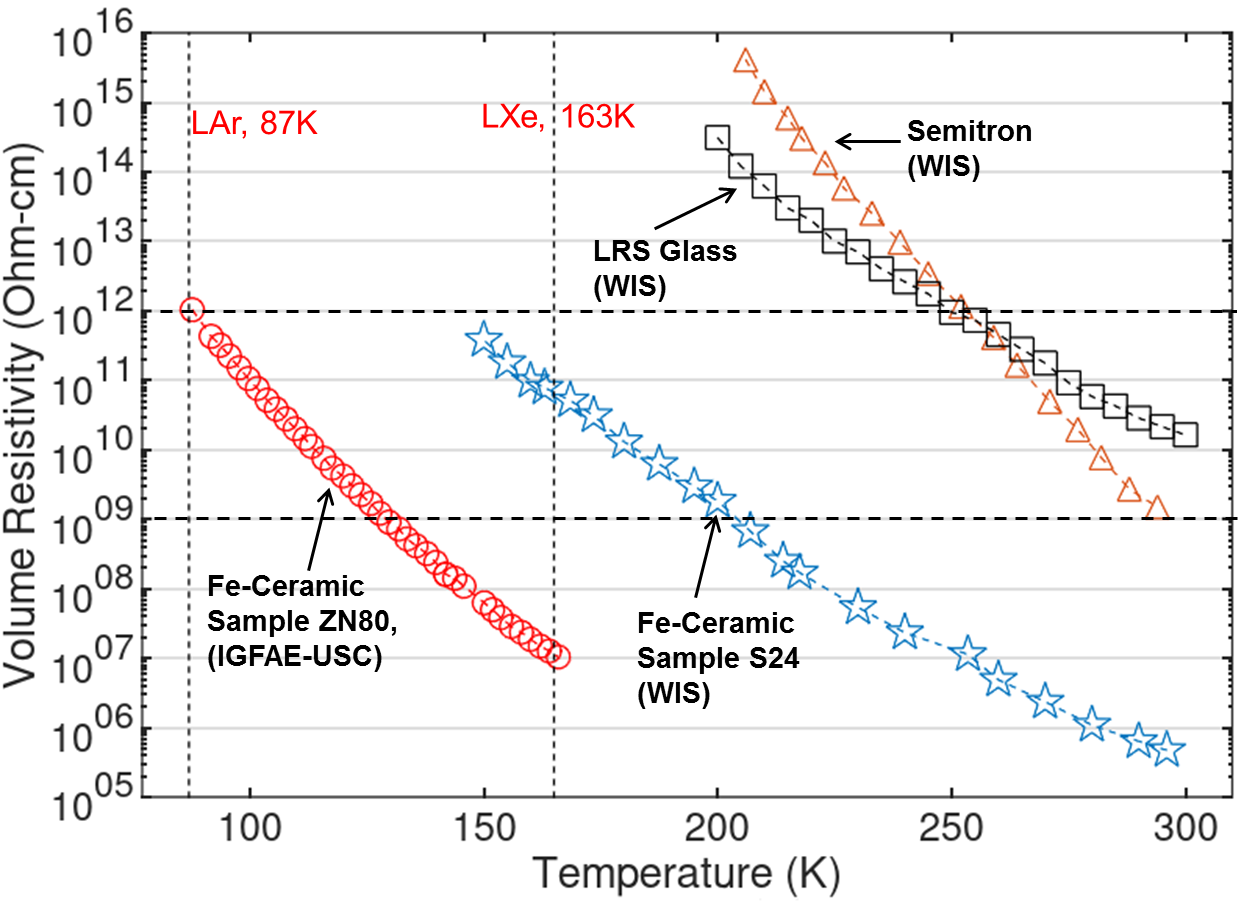}
\caption{\label{fig:ii}Temperature dependence of the volume resistivity for Semitron, LRS glass and two samples of ferrite ceramics. While the semitron and the LRS glass effectively become insulators below $\sim$ 200 K, the Fe-ceramic samples S24 and ZN80 can have their resistivity tuned to suit RPWELL operation at LXe and LAr temperature, respectively. }
\end{figure}

The resistive plates currently used for RT operation are Semitron ESD 225, which is an acetal-based, static dissipative plastic with $\rho$ $\sim$$10^{9}$ $\Omega$$\cdot$cm) (Quadrant Plastics, USA) and Low Resistive Silicate (LRS) glass with $\rho$ $\sim$2 $\times$ $10^{10}$ $\Omega$$\cdot$cm (Tsinghua University, Beijing, China). Both Semitron and LRS glass quench discharges effectively at RT, as demonstrated in laboratory and accelerator experiments with RPWELL detectors operated at gains of a few times $10^{4}$,  in different gas mixtures and in high-rate radiation environments \cite{25,26,28}. The measurement of resistivity as a function of temperature performed with Semitron and LRS glass indicated an Arrhenius behavior, i.e. exponential increase in resistivity with decreasing T. As shown in Figure \ref{fig:ii}, the Semitron and LRS glass plates have resistivity values $\sim$$10^{14}$ $\Omega$$\cdot$cm at 200 K; they behave as insulators at lower temperatures, rendering them unsuitable for RPWELL operation in LAr and LXe.

One of the remarkable advantages of the robust Fe-Ceramics \cite{42}, developed for Resistive-Plate Chambers, are their tunable electrical properties. The ability to tune their electric resistivity by post-processing treatments at 500 to 800$^{\circ}$C, to be within the range $\sim$10$^{9}$ - 10$^{12}$ $\Omega$$\cdot$cm at LXe and LAr temperatures, makes them an excellent candidate as a resistive plate for cryogenic RPWELL and other resistive gas-avalanche detectors. The temperature dependence of the resistivity of the Fe-ceramics, shown in Figure \ref{fig:ii}, indicates that starting with a resistivity of $\sim$$10^{7}$ $\Omega$$\cdot$cm at RT, the latter increases exponentially to $\sim$$10^{11}$ $\Omega$$\cdot$cm at 163 K.
Similarly, tuning the resistivity to $\sim$$10^{5}$ $\Omega$$\cdot$cm at room temperature leads to $\sim$$10^{12}$ $\Omega$$\cdot$cm at LAr temperature (87 K). A list of other materials investigated that failed to quench discharges at cryogenic temperatures are highlighted in Table \ref{tab:my-table}.

\begin{table}[]
\caption{List of materials failing to quench discharges at cryogenic temperatures}
\label{tab:my-table}
\begin{tabular}{|c|c|c|c|c|}
\hline
\textbf{Material}                                                                          & \textbf{Source}                                                        & \textbf{Resistivity @ RT}                                                                 & \textbf{\begin{tabular}[c]{@{}c@{}}Resistivity \\ as function of T\end{tabular}}           & \textbf{\begin{tabular}[c]{@{}c@{}}Discharge \\ Quenching at Cryo T\end{tabular}} \\ \hline
\begin{tabular}[c]{@{}c@{}}Tivar EC \\ \& \\ Tivar ESD\\ (UHMW-PE)\end{tabular}            & \begin{tabular}[c]{@{}c@{}}J. Vavra\\ SLAC, USA\end{tabular}        & \begin{tabular}[c]{@{}c@{}}$\rho$ $\sim$ $10^{6}$ \\ - $10^{7}$ $\Omega$$\cdot$cm\end{tabular}  & Constant                                                                                   & \begin{tabular}[c]{@{}c@{}}Fails; \\ low $\rho$ value\end{tabular}                \\ \hline
\begin{tabular}[c]{@{}c@{}}PTFE\\  + 1.5\% Carbon\end{tabular}                             & 3M, USA                                                                & \begin{tabular}[c]{@{}c@{}}$\rho$ $\sim$ $10^{7}$ \\ - $10^{8}$ $\Omega$$\cdot$cm\end{tabular}  & Constant                                                                                   & \begin{tabular}[c]{@{}c@{}}Fails;\\ low $\rho$ value\end{tabular}                 \\ \hline
\begin{tabular}[c]{@{}c@{}}Araldite \\ + Graphite \\ (Graphite \\ - 15-30 \%)\end{tabular} & \begin{tabular}[c]{@{}c@{}}Fabricated \\ @WIS\end{tabular}             & \begin{tabular}[c]{@{}c@{}}$\rho$ $\sim$ $10^{8}$ \\ - $10^{14}$ $\Omega$$\cdot$cm\end{tabular} & Constant                                                                                   & \begin{tabular}[c]{@{}c@{}}Fails;\\ non-uniform \\ resistivity\end{tabular}       \\ \hline
\begin{tabular}[c]{@{}c@{}}Si-based \\ Ceramics\end{tabular}                               & \begin{tabular}[c]{@{}c@{}}L. Naumann,\\ HZDR, \\ Germany\end{tabular} & $\rho$ $\sim$ $10^{8}$ $\Omega$$\cdot$cm                                                        & \begin{tabular}[c]{@{}c@{}}Increases \\ exponentially\\  with \\ decreasing T\end{tabular} & \begin{tabular}[c]{@{}c@{}}Fails;\\ $\rho$ too high \\ @ LXe T\end{tabular}       \\ \hline
\end{tabular}
\end{table}


\section{Experimental setup and methodology}

\begin{figure}[htbp]
\centering 
\includegraphics[width=.6\textwidth]{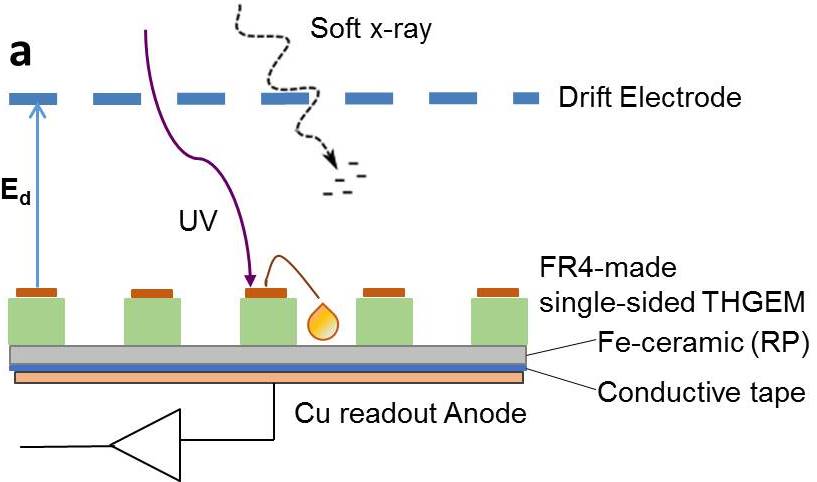}
\qquad
\includegraphics[width=.6\textwidth]{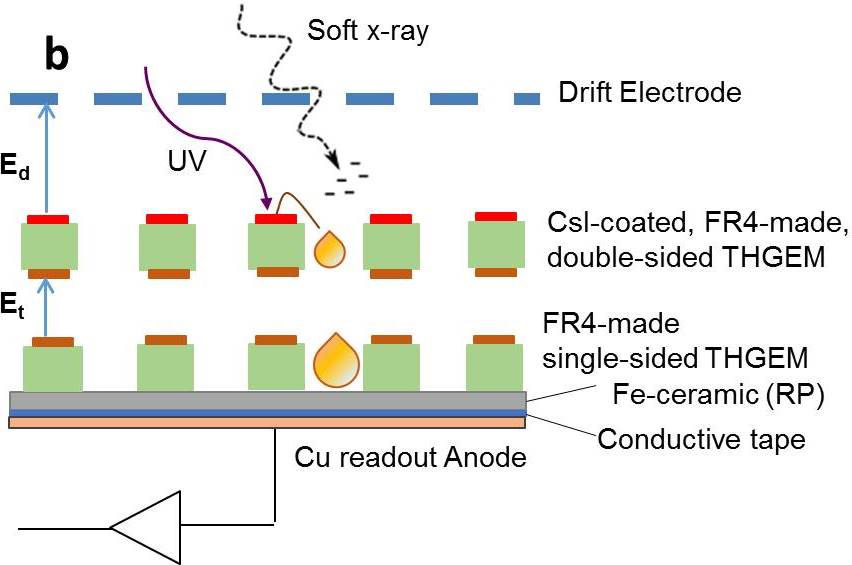}
\caption{\label{fig:iii}a) A schematic diagram of the single-stage RPWELL detector, a single-sided THGEM pressed against a Fe-ceramic resistive plate ($\rho$ $\sim$$10^{7}$ $\Omega$$\cdot$cm at RT and $\sim$$10^{11}$ $\Omega$$\cdot$cm at 163 K), glued with double-sided conductive Ag tape to a Cu readout anode. As a reference, the same single-sided THGEM was tested in the Thick-WELL (THWELL) configuration, coupling it directly to the metallic anode. b) A schematic diagram of the double-stage cascaded THGEM and Fe-ceramic RPWELL detector; charges preamplified in the THGEM are transferred with the transfer field $E_t$, and amplified in the RPWELL.}
\end{figure}

Schematic diagrams of the single and double-stage detectors investigated in this work are shown in Figure  \ref{fig:iii}. The single-stage RPWELL detector (Figure \ref{fig:iii}a) comprised a single-sided THGEM, made of FR4 (glass-reinforced epoxy laminate), with thickness $t = 0.6$ mm , hole diameter $d = 0.5$ mm, hole pitch $a = 1$ mm, and an etched rim around each hole of width, $h = 0.1$ mm. The THGEM electrode was pressed against a Fe-ceramic resistive plate. The volume resistivity of the ceramic plate (Sample S24 of Figure \ref{fig:ii}) was measured to be $\sim$$10^{7}$ $\Omega$$\cdot$cm and $\sim$$10^{11}$ $\Omega$$\cdot$cm at RT and at 163 K, respectively (Figure \ref{fig:ii}). The resistive plate was glued to the Cu anode with a double-sided conductive Ag tape (Electrically Conductive Adhesive Transfer Tape 9707, 3M, USA). A metallic mesh (drift electrode) placed at a distance of 5 mm from the THGEM top was used to set the drift field, $E_d$. The detector assembly was mounted in a sealed stainless-steel chamber; it was cooled by immersing it in an ethanol dewar cooled by  LN$_{2}$. The procedure involved the following steps: 
\noindent pumping the detector chamber down to $10^{-5}$ mbar; flushing with Ne/5$\%$CH$_{4}$ at 1 bar, at RT for 3 hours at a flow of 50 sccm; sealing the chamber and slowly immersing it in the dewar (additional LN$ _{2}$ was added for further cooling). 

The operation in sealed mode ensured stable thermodynamic conditions during measurements. Voltages were ramped up and data were taken only after the temperature (measured with a thermocouple touching the resistive plate) stabilized at 163 K (at a pressure of $\sim$800 mbar).  Measurements with 5.9 keV X-rays were done using a collimated $^{55}$Fe source placed inside the chamber, illuminating the detector with a $\sim$2 mm diameter beam spot. Measurements with single UV-photons were performed by shining UV light from a continuously emitting Ar(Hg) lamp through a fused silica window at the top of the cryogenic setup. The light was passed through an optical filter of optical density 3.0 and a 0.1 mm pinhole to reduce the rate down to $\sim$100 Hz. HV bias was supplied to the THGEM (top and bottom) and RPWELL electrodes using CAEN N471A power-supplies, via low-pass filters. The currents from the power-supply channels (monitor output) were recorded with a National Instruments 6008 Data Acquisition I/O data logging device using the National Instruments Signal Express software. The recording of the current drawn by the electrodes throughout the experiment permitted monitoring the discharge behavior of the detector. The avalanche-charge signals induced on the  readout anode were recorded with a Canberra 2006 charge-sensitive preamplifier and shaped by an ORTEC 571 linear amplifier (2 $\mu$s shaping time). Pulse-height spectra were acquired with a multichannel analyzer (MCA), model AMPTEK MCA-8000D, using the unipolar signals from the linear amplifier. The entire electronics chain was calibrated using a pulse generator and a known input capacitor. As a reference, measurements with x-rays were carried out under similar conditions, with the same single-sided THGEM -- directly coupled to the Cu anode, in a Thick-WELL (THWELL \cite{8}) configuration. Gain-curves were measured at a drift field ($E_d$) of 0.5 kV/cm throughout the experiment.

\section{Results}

\subsection {Operation of a single-stage Fe-ceramic RPWELL at 163 K}

\begin{figure}[!htbp]
\centering
\includegraphics[width=.8\textwidth]{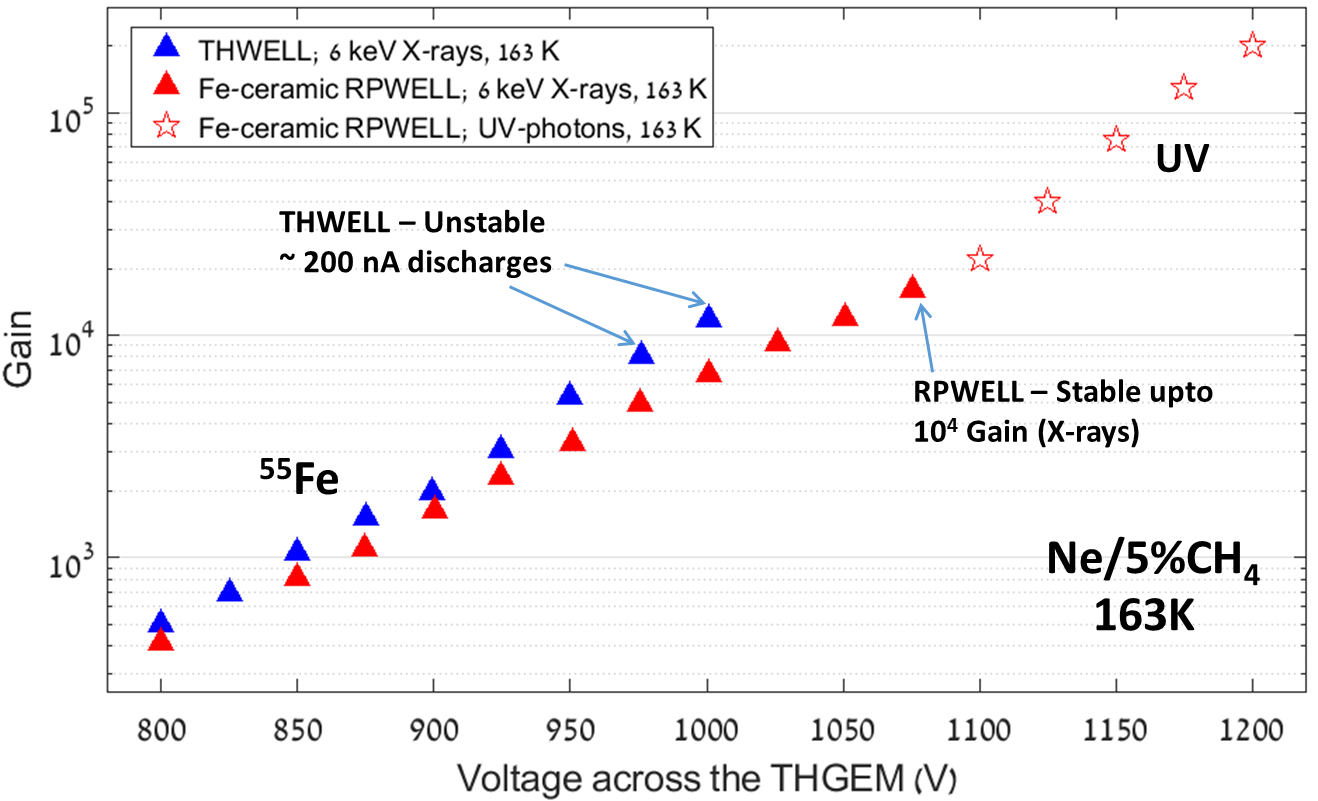}
\caption{\label{fig:iv}Gain curves vs voltage across the electrode, of the single-stage Fe-ceramic RPWELL (of Figure \ref{fig:iii}a) with 5.9 keV X-rays and single UV-photons in a sealed chamber with Ne/5$\%$CH$_{4}$ at 163 K and 800 mbar; $E_d$ = 0.5 kV/cm. The gain curve of a single-stage THWELL was measured for comparison in the same setup. The RPWELL was operated here without evaporating CsI on the THGEM electrode.}
\end{figure}

\begin{figure}[!ht]
\centering
\includegraphics[width=.96\textwidth]{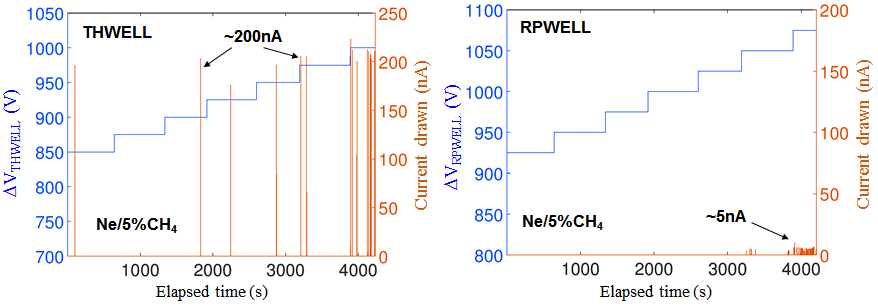}
\caption{\label{fig:v} Discharge behavior of the single-stage THWELL and RPWELL detectors as a function of time, demonstrating the effect of the resistive plate. While $\sim$200 nA high discharges occurred during the THWELL operation above 850 V (gain $\sim$$10^{3}$; Figure \ref{fig:iv}), they are quenched by the resistive plate in the RPWELL; the $\sim$5 nA high discharges appear at gains$>$$10^{4}$.}
\end{figure}

The single-stage RPWELL (Figure \ref{fig:iii}a) with the Fe-ceramic resistive plate (sample S24 of Figure \ref{fig:ii}) was investigated with 5.9 keV X-rays and single UV-photons in a sealed chamber with Ne/5$\%$CH$_{4}$ at 163 K and 800 mbar; $E_d$ was kept at 0.5 kV/cm. This is the first demonstration of RPWELL-detector operation at cryogenic conditions. The performance of a single-stage THWELL (with the same THGEM electrode coupled directly to the Cu anode) was studied in identical conditions, as a reference. The results of the comparative gain measurements are shown in Figure  \ref{fig:iv}. The RPWELL operated discharge-free up to a gain $\sim$$10^{4}$ with X-rays and $\sim$$10^{5}$ with single UV-photons (without CsI photocathode, i.e. with photoelectrons emitted from the THGEM-top Cu surface). The maximum achievable gain with the single-stage detectors irradiated with X-rays in Ne/5$\%$CH$_{4}$ gas mixtures ($\sim$200 primary electrons) was observed to be $\sim$10-fold lower compared to that with single photoelectrons. This can be attributed to charge spreading between several holes in the case of X-rays, relaxing the constraint imposed by the theoretical Raether limit \cite{4}. Discharges with X-rays appeared at higher gain in the RPWELL and were less intense as shown in Figure  \ref{fig:v} -- a clear manifestation of the discharge quenching property of the resistive plate. The THWELL operation at high gain was unstable due to regular $\sim$200 nA discharges, starting already at low gain ($\sim$$10^{3}$ at 850 V). In comparison, the RPWELL detector had occasional $\sim$5 nA discharges, starting at gains $>$$10^{4}$.

\begin{figure}[!htbp]
\centering
\includegraphics[width=.93\textwidth]{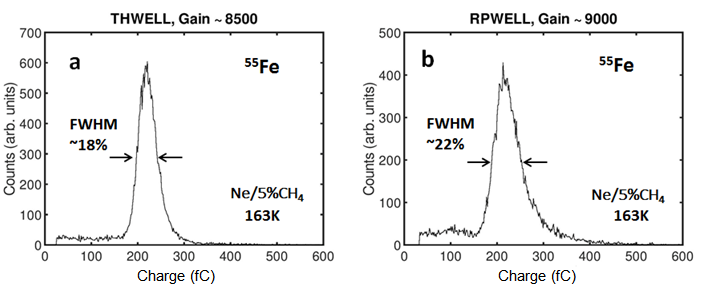}
\caption{\label{fig:vi} Pulse-height spectra of the a) THWELL and b) RPWELL detectors measured with uncollimated 5.9 keV X-rays in Ne/5$\%$CH$_{4}$ at 163 K and 800 mbar, $E_d$ = 0.5 kV/cm, at gains of $\sim$$10^{4}$. }
\end{figure}

The pulse-height spectra with uncollimated  X-rays recorded in the THWELL and the RPWELL detectors at 163 K, in identical conditions, are shown in Figure \ref{fig:vi}. The detectors operated in Ne/5$\%$CH$_{4}$ at gains of $\sim$$10^{4}$. The THWELL detector was unstable and suffered occasional discharges at this gain. The THWELL yielded an energy resolution of $\sim$18$\%$ (FWHM) while the RPWELL yielded (in this set on measurements) a slightly broader resolution $\sim$22$\%$ (FWHM). The energy resolution observed with the Fe-ceramic RPWELL is not fully understood yet. A possible explanation could be the non-uniformity of the Fe-ceramic samples due to an inadequate surface polish that could result in degradation of the energy resolution when the avalanche spreads over larger areas. Detailed investigations are ongoing to understand this behavior.

\subsection {Double-stage THGEM and Fe-ceramic RPWELL operation at 163 K}

\begin{figure}[!htbp]
\centering
\includegraphics[width=.9\textwidth]{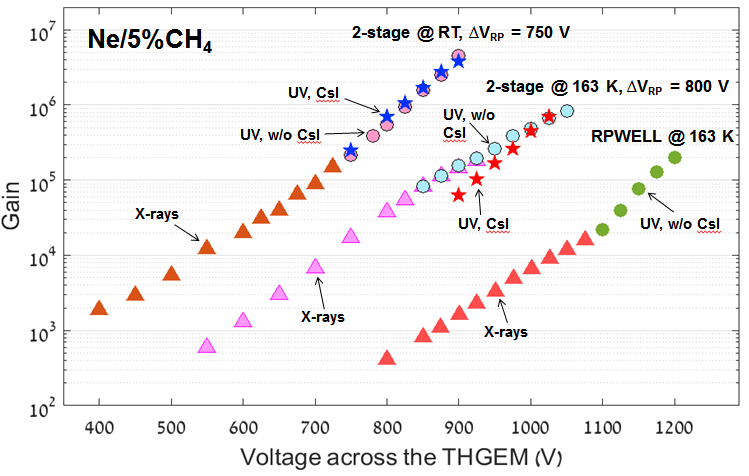}
\caption{\label{fig:vii} Gain curves of the cascaded THGEM + RPWELL (2-stage) detector (of Figure \ref{fig:iii}b) with X-rays and single UV-photons at RT (1000 mbar) and 163 K (800 mbar) in Ne/5$\%$CH$_{4}$, with and without a CsI photocathode. The double-stage gain measurements were performed as a function of the THGEM voltage (at $E_d$ = 0.5 kV/cm for X-rays and $E_d$ = 0 for UV-photons ; $E_t$ = 0.5 kV/cm), keeping the RPWELL voltage constant ($\Delta V_{RP} = 750$ V at RT and $\Delta V_{RP} = 800$ V at 163 K), until the appearance of occasional discharges on the THGEM. The single-stage RPWELL gain curves are shown here as reference.}
\end{figure}

The operation of the double-stage detector aimed at reaching higher stable gains. The gain curves of the THGEM + RPWELL detector, shown in Figure \ref{fig:vii} (at RT and 163 K), were recorded at a constant RPWELL voltage ($\Delta V_{RP} = 750$ V at RT and $\Delta V_{RP} = 800$ V at 163 K), while increasing that of the THGEM, until the appearance of ocassional discharges in the latter. As seen in Figure \ref{fig:vii}, the double-stage detector reached 10-fold higher gains than the single-stage RPWELL, namely $\sim$$10^{5}$ with X-rays at RT. Measurements with single UV-photons at RT were performed with and without CsI-coating on the top-THGEM electrode surface; in both cases the double-stage detector reached gains $\sim$5 $\times$ $10^{6}$. During double-stage detector operation at RT, the RPWELL detector had occasional $\sim$25 - 30 nA discharges at gains $>$$10^{5}$, as the Fe-ceramic plate fails to quench discharges completely due to its low resistivity ($\rho$ $\sim$$10^{7}$ $\Omega$$\cdot$cm).

The operation of the double-stage detector at 163 K resulted in gains $\sim$$10^{5}$ with X-rays. The measurement at 163 K with single UV-photons yielded stable gains of $\sim$7 $\times$ $10^{5}$ and $\sim$8.5 $\times$ $10^{5}$, with and without CsI-coating on the top-THGEM surface, respectively. There were no discharges observed on the RPWELL and the measurements were stopped due to the appearance of occasional discharges on the THGEM (at gains beyond the maximum achievable stable gains mentioned above). The temperature gradients inside the detector chamber lead to density gradients, such that the gas density is higher in the avalanche region at 163 K than at RT and this seems to be the most likely explanation behind the lower maximum achievable gain obtained with the double-stage detector at 163 K.

\begin{figure}[!htbp]
\centering
\includegraphics[width=0.98\textwidth]{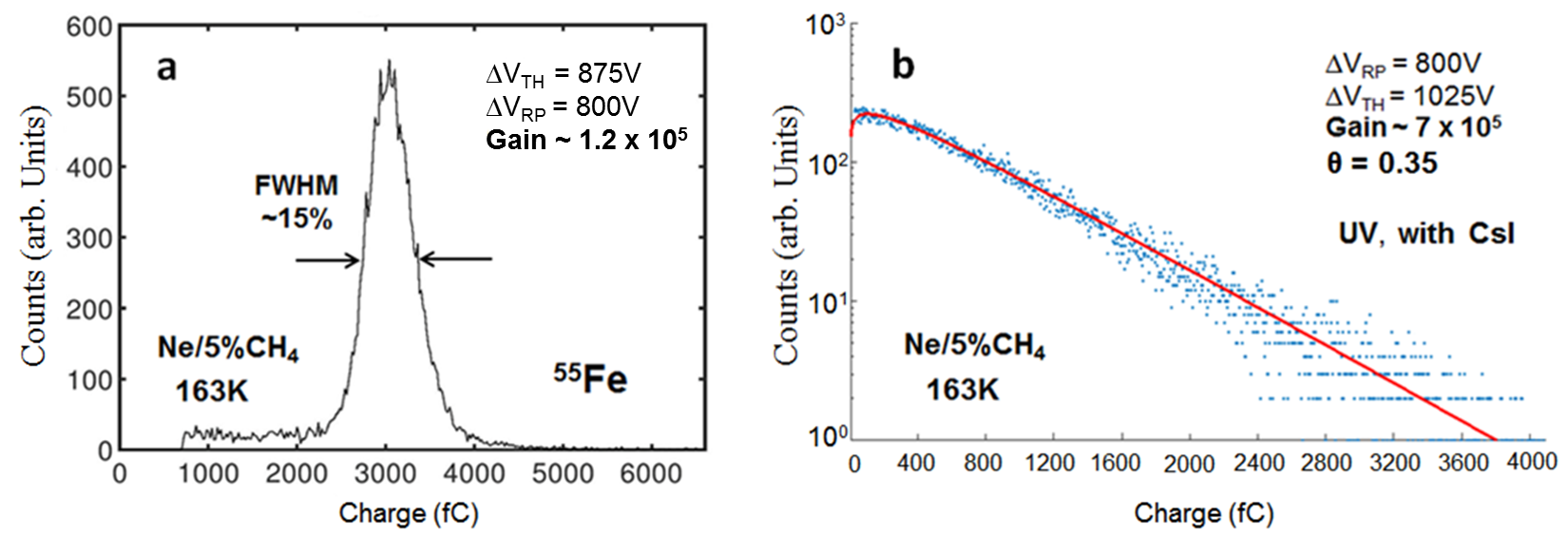}
\caption{\label{fig:viii}The pulse-height spectra of the double-stage detector (of Figure \ref{fig:iii}b) operated at 163 K in 800 mbar Ne/5$\%$CH$_{4}$ : a) with collimated (1mm collimator) 5.9 keV X-rays at a gain of $\sim$$10^{5}$, at a THGEM voltage  -- $\Delta V_{TH} = 875$ V and an RPWELL voltage -- $\Delta V_{RP} = 800$ V; b)  with single UV-photons at a gain of $\sim$$10^{6}$, at a THGEM voltage -- $\Delta V_{TH} = 1025$ V and an RPWELL voltage -- $\Delta V_{RP} = 800$ V, showing a clear Polya distribution.}
\end{figure}

The pulse-height spectra of the double-stage detector operated at 163 K  in 800 mbar Ne/5$\%$CH$_{4}$ are shown in Figure \ref{fig:viii}, for 5.9 keV X-rays (detector gain $\sim$$10^{5}$) and single UV-photons (detector gain $\sim$7$\times$$10^{5}$). The energy resolution measured with 5.9 keV X-rays (collimated to 1 mm diameter) is $\sim$15$\%$ FWHM. The UV-induced spectrum has a clear Polya distribution and the Polya-fitted spectrum at $\sim$7$\times$$10^{5}$ gain has a $\theta$ $\sim$ 0.35. The single-electron detection efficiency estimated numerically from the fitted spectrum is $\sim$90$\%$ at a moderate electronics threshold $\sim$$10^{5}$ electrons.

\section{Summary and discussion}

The availability of ferric-based (Fe$^{3+}$) ceramic composite (Fe-ceramic) resistive materials, tuned to a volume resistivity in the range of $\sim$10$^{9}$ - 10$^{12}$ $\Omega$$\cdot$cm, permitted, for the first time, operating an RPWELL detector at temperatures down to 163 K (LXe temperature, close to the Xe triple point). Other materials investigated here in this context, e.g. Semitron and LRS glass (adequate for RT operation) failed, due to exponentially increased resistivity at low temperatures.  Two detector configurations were investigated : a single-stage RPWELL and a double-stage RPWELL preceded by a THGEM pre-amplifying element. The detectors were investigated at 163 K, in 800 mbar Ne/5$\%$CH$_{4}$, with a Fe-ceramic plate of resistivity $\sim$$10^{11}$ $\Omega$$\cdot$cm. Detailed investigations to tune the resistivity of new Fe-ceramic samples for operation at LAr temperature (87 K) are ongoing.

The single-stage RPWELL exhibited discharge-free operation at 163 K, up to gains of $\sim$$10^{4}$ with X-rays and $\sim$$10^{5}$ with single UV-photons. Comparison with a bare THWELL demonstrated the effect of the resistive plate in quenching discharges. The THWELL encountered $\sim$200 nA discharges starting at gains $\sim$$10^{3}$ with X-rays, with increased frequency at higher gains -- leading to unstable operation. In contrast, the onset of $\sim$forty-fold smaller, 5 nA discharges, started in the RPWELL at gains above $10^{4}$ with X-rays. The energy resolution measured with uncollimated 5.9 keV X-rays with the THWELL and the single-stage RPWELL were $\sim$18 $\%$ and $\sim$22$\%$ FWHM respectively. The energy resolution observed with the Fe-ceramic RPWELL is possibly due to a relatively coarse polishing of the first Fe-ceramic samples, and is currently under investigation.

A double-stage cascaded THGEM and Fe-ceramic RPWELL detector yielded discharge-free operation at 163 K in  800 mbar Ne/5$\%$CH$_{4}$, at about ten-fold higher gains compared to the single-stage : $\sim$$10^{5}$ with X-rays and with single UV-photons gains $\sim$7 $\times$ $10^{5}$ and $\sim$8.5 $\times$ $10^{5}$ were recorded, with and without a CsI photocathode, respectively. The energy resolution measured with 5.9 keV x-rays (collimated to 1 mm diameter) was $\sim$15$\%$ FWHM. The UV-photons yielded clear Polya distributions -- relevant for enhancing the detection efficiency of single photons, e.g. in cryogenic gas-avalanche photomultipliers (GPM). 

The successful operation of RPWELL detectors at cryogenic temperatures constitutes an important milestone towards their potential applications in future large-volume noble liquid (LXe and LAr) detectors in neutrino physics, dark-matter searches and other basic and applied fields.

\acknowledgments
This research work was partially funded by the Israel Science Foundation, grant number 1719/16. We thank Dr. J. Vavra (SLAC) and Dr. L. Naumann (HZDR), for supplying the resistive-plate samples. DGD acknowledges the Ramon y Cajal program (Spain) under contract number RYC-2015-18820. This research work was performed within the framework of the CERN RD51 collaboration.

\end{document}